\documentclass{PoS}
\usepackage{cite}
\usepackage{mcite}
\usepackage{array}
\usepackage{epsfig}
\usepackage{amssymb}
\usepackage{graphics,graphpap}
\usepackage{amsmath}
\usepackage{slashed}
\usepackage{url}
\usepackage{booktabs}

\newcommand{\be}{\begin{equation}}
\newcommand{\ee}{\end{equation}}
\newcommand{\bea}{\begin{eqnarray}}
\newcommand{\eea}{\end{eqnarray}}

\newcommand{\bi}{\begin{itemize}}
\newcommand{\ei}{\end{itemize}}

\newcommand{\neu}[1]{\tilde{\chi}^0_{#1}}
\newcommand{\cha}{\tilde{\chi}}

\newcommand{\mneu}[1]{m_{\tilde{\chi}^0_{#1}}}
\newcommand{\mcha}[1]{m_{\tilde{\chi}^\pm_{#1}}}

\newcommand{\gev}{\ \mathrm{GeV}}

\newcommand\T{\rule{0pt}{2.5 ex}}

\newcommand\B{\rule[-1.5ex]{0pt}{0pt}}

\title{Determining MSSM parameters via chargino production at the LC: a one-loop analysis}

\ShortTitle{Determining MSSM parameters via chargino production at the LC: a one-loop analysis}

\author{\speaker{Aoife Bharucha}\\
        University of Hamburg\\
        E-mail: \email{aoife.bharucha@desy.de}}
\abstract{Very precise measurements of masses and cross sections are expected to be achievable 
with a future linear collider. With such an accuracy one must incorporate loop corrections in 
order to make meaningful predictions for the underlying 
new physics parameters. For the electroweakino sector, this involves fitting one-loop 
predictions to expected measurements of the cross section and forward-backward asymmetry for 
chargino pair production and of the accessible chargino and neutralino masses. We consider two 
scenarios with characteristic features, chosen taking recent LHC SUSY and Higgs searches 
into account. Our analysis allows the accurate determination of the desired parameters and, 
additionally, access to stop sector parameters that enter via loop corrections.}

\FullConference{36th International Conference on High Energy Physics,\\
		July 4-11, 2012\\
		Melbourne, Australia}

\begin{document}

\section{Introduction}\label{sec:1}

A linear collider (LC) is ideally suited to high precision studies of physics beyond the standard model (BSM).
Light electroweakinos, motivated by naturalness arguments~ref.~\cite{Hall:2011aa} and 
GUT motivated SUSY models~\cite{Brummer:2011yd}, could well be within reach of a first stage LC.
This would allow the nature of the underlying SUSY model to be probed via the determination of the
fundamental parameters.
LO studies (see
ref.~\cite{Desch:2003vw} and references therein) have shown that the U(1) parameter $M_1$, the SU(2) parameter $M_2$, the
higgsino 
parameter $\mu$ and $\tan\beta$, the ratio of the vacuum expectation values of
the two neutral 
Higgs doublet fields, can be determined at the
percent level via chargino and neutralino pair-production.
However, as one-loop effects in the MSSM can be large, higher order calculations are crucial in order to perform an accurate assessment.
These corrections mean that additional MSSM parameters become relevant, such as the masses of the stops which are so far 
only weakly constrained by the LHC.
Fitting experimental results to loop corrected predictions, calculated in the on-shell scheme, it should therefore be possible to 
extract the parameters of the chargino--neutralino sector of the MSSM Lagrangian, as well as to indirectly gain insight 
into other sectors.
In sec.~\ref{sec:one-loop} we discuss the strategy used to fit the MSSM parameters and briefly 
describe the calculation of NLO corrections. Then in sec.~\ref{sec:num-res} we present our numerical results and
finally in sec.~\ref{sec:conc} we discuss the implications of these results.

\section{Fit strategy and the calculation of NLO corrections}\label{sec:one-loop}
In the chargino and neutralino sectors there are four real parameters which we fit to: $M_1$, $M_2$, $\mu$ and $\tan\beta$.
If beyond the direct reach of the LC, the sneutrino mass is also included in the fit. Depending on the scenario, only limited
knowledge about the remaining MSSM parameters, e.g. for the stop sector, which will all contribute at the loop level, may be available. 
Our analysis also offers unique indirect sensitivity to such parameters at the
LC. In the fit we use the polarised cross-sections and forward backward asymmetry for chargino
production as well as the $\cha_1^{\pm},\cha_2^{\pm}$
and $\neu{1}, \neu{2}, \neu{3}$ masses, calculated at NLO in an on-shell scheme as described below.
We assume the masses to have been measured at the LC using the threshold scan method, however we also investigate how the
fit precision changes if the masses were obtained from the continuum~\cite{AguilarSaavedra:2001rg}.
Further details of the fit method and errors are given in ref.~\cite{Bharucha:2012ya}.
We carry out the fit for two scenarios, S1 and S2, shown 
in tab.~\ref{tab:s1}.\footnote{Note that S2 corresponds to S3 in ref.~\cite{Bharucha:2012ya}.}
Due to the current status of direct LHC searches~\cite{atlas:2012rz,*cms:2012jx},
in both scenarios we require first and second generation squarks and
gluinos to be heavy, but on the other hand assume the stop sector to be light.
Further, indirect limits, checked using 
{\texttt{micrOmegas 2.4.1}}~\cite{Belanger:2006is, *Belanger:2010gh}, lead us to choosing mixed gaugino higgsino scenarios, 
favoured by the relic density measurements~\cite{Komatsu:2010fb}
and relatively high pseudoscalar Higgs masses, preferred by flavour physics 
constraints i.e.\ the branching ratio $\mathcal{B}(b\to s\gamma)$ 
and $\Delta(g_\mu-2)/2$.
Our scenarios are chosen in order to assess the fit sensitivity for different possible 
choices of parameters e.g. the pseudoscalar Higgs mass $M_{A^0}$.
Note that for S2, $m_h$ (=125 GeV) is compatible
with the recent Higgs results from the LHC~\cite{ATLAS:2012gk,*CMS:2012gu}.

The one-loop corrections to the polarised cross-section and forward backward asymmetry for
$e^+e^-\to \tilde{\chi}^+_1\tilde{\chi}^-_1$ are calculated in full within the MSSM, following~\cite{bfmw,*Bharucha:2012re}, using the program
\texttt{FeynArts}~\cite{Denner:1992vza},
which allowed an automated generation of the Feynman diagrams and
amplitudes. Together with the packages 
\texttt{FormCalc}~\cite{FormCalc2} and 
\texttt{LoopTools}~\cite{Hahn:1998yk} we derived
the final matrix elements and loop integrals.
The regularisation method we use is dimensional
reduction (see e.g.~\cite{0503129}), which ensures that SUSY is 
preserved, as in e.g. 
ref.~\cite{delAguila:1998nd}. The results at NLO depend on many MSSM parameters beyond the small subset at tree level.
Finite results at one-loop order are obtained by adding counterterm diagrams, also generated by \texttt{FeynArts}, however expressions 
for the counterterms which renormalise the tree-level couplings are required as input, as given in detail
in ref.~\cite{Bharucha:2012ya}.
Using this prescription to renormalise the vertices we obtain UV-finite results.
Soft and collinear radiation must be included to obtain a result free of infra-red and collinear singularities.
In the regions $E<\Delta E$ and $\theta<\Delta \theta$ where $\Delta E$ and $\Delta \theta$ denote the cut-offs, the radiative cross-section 
can be factorised into analytically integrable expressions proportional to the tree-level cross-section 
$\sigma^{\rm tree}(e^+e^-\to \tilde{\chi}_i^+\tilde{\chi}_j^-)$. 
The soft contribution can easily be incorporated using \texttt{FormCalc}, however the collinear contribution must be added explicitly,
this is done as described in ref.~\cite{Bharucha:2012ya}.

\section{Numerical results}\label{sec:num-res}
\begin{table}[tb!]
\renewcommand{\arraystretch}{1.1}
\begin{center}
 \begin{tabular}{lclclclc}
\toprule
\multicolumn{4}{c}{\footnotesize{S1}}& \multicolumn{4}{c}{\footnotesize{S2}}\\ 
\vspace{-0.9cm} \\
\multicolumn{4}{c}{\downbracefill}&\multicolumn{4}{c}{\downbracefill}\\
\T Parameter & Value & Parameter & Value& Parameter & Value & Parameter & Value\\
\midrule
\T $M_1$ & 125& $M_2$ & 250&$M_1$ & 106& $M_2$ & 212\\
$\mu$ & 180 & $M_{A^0}$ & 1000&$\mu$ & 180 & $M_{A^0}$ & 500\\
$M_3$ & 700 & $\tan\beta$ & 10&$M_3$ & 1500 & $\tan\beta$ & 12\\
$M_{e_{1,2}}$ & 1500 & $M_{e_{3}}$ & 1500&$M_{e_{1,2}}$ & 125 & $M_{e_{3}}$ & 106\\
$M_{l_{i}}$ & 1500 & $M_{q_{1,2}}$ & 1500 &$M_{l_{i}}$ & 180 & $M_{q_{i}}$ & 1500 \\
\B $M_{{q/u}_{3}}$ & 400 & $A_f$ & 650&$M_{{u}_{3}}$ & 450 & $A_f$ & -1850\\
\bottomrule
\end{tabular}
\caption{Table of parameters (with the exception of $\tan\beta$ in GeV), for scenarios 1 (S1) and 2 (S2). Here $M_{(l/q)_{i}}$ and
$M_{(e/u)_{i}}$ represent the left and right handed mass parameters for of a
slepton/squark of generation $i$ respectively, and $A_f$ is the trilinear
coupling for a sfermion $\tilde f$.\label{tab:s1}}
\end{center}
\end{table}
In S1, only the charginos and three neutralinos will be accessible at the
LC. It should be possible to probe the supersymmetric QCD sector, 
with masses of $\sim$1.5~TeV, at the LHC. As input for the fit we
therefore use: the masses of the charginos ($\cha_1^{\pm},\,\cha_2^{\pm}$) and three lightest
neutralinos ($\neu{1},\,\neu{2},\,\neu{3}$); the production cross section $\sigma(\cha^+_1\cha^-_1)$
with polarised beams at $\sqrt{s} = 350$ and $500 \gev$; the forward-backward asymmetry $A_{FB}$ at $\sqrt{s} = 350$ and $500 \gev$; 
the branching ratio $\mathcal{B}(b\to s\gamma)$ calculated using
{\texttt{micrOmegas 2.4.1}}~\cite{Belanger:2006is, *Belanger:2010gh}, see tab.~\ref{tab:inputsc1}.
Note that $\mathcal{B}(b\to s\gamma)$ increases sensitivity to 
the third generation squark sector, and the estimated experimental precision is $0.3\cdot 10^{-4}$~\cite{O'Leary:2010af}. 
As the impact of the muon anomalous magnetic moment is negligible it is not included in the fit.

For S2, the input for the fit, as summarised in tab.~\ref{tab:inputsc1}, is the same as in S1, however we consider a centre of mass energy $\sqrt{s}=400\gev$ instead of $350\gev$,
and this is supplemented by the Higgs
boson mass, $m_h$, with a theoretical uncertainty of $\sim$1 GeV, calculated using \texttt{FeynHiggs
2.9.1}~\cite{Heinemeyer:1998yj,*Degrassi:2002fi,*Frank:2006yh}. 
The estimated experimental precision at the LC for $m_h$, taken from
ref.~\cite{AguilarSaavedra:2001rg}, is adopted. The anomalous muon magnetic moment $\Delta (g_\mu -2)/2$, calculated using~{\texttt{micrOmegas 2.4.1}}~\cite{Belanger:2006is,
*Belanger:2010gh}, plays a role and is therefore included in the fit, with a projected  
experimental error of $3.4\cdot10^{-10}$~\cite{Carey:2009zzb}, which we assume would dominate
over the theoretical uncertainty. 
We assume that the sneutrino mass has been measured. 
\begin{table}[tb!]
\renewcommand{\arraystretch}{1.1}
%\vspace{1cm}
\begin{center}
\begin{tabular}{@{}lr@{.}lr@{.}lr@{.}lr@{.}l@{}r@{.}lr@{.}lr@{.}lr@{}l@{}}
\toprule
&\multicolumn{8}{c}{\footnotesize{S1}}& \multicolumn{8}{c}{\footnotesize{S2}}\\ 
\vspace{-0.9cm} \\&\multicolumn{8}{c}{\downbracefill}&\multicolumn{8}{c}{\downbracefill}\\
\T Observable & \multicolumn{2}{c}{Tree} & \multicolumn{2}{c}{\quad Loop} & \multicolumn{2}{c}{$\delta_{\rm exp}$} & \multicolumn{2}{c}{$\delta_{\rm th}$}& \multicolumn{2}{c}{\quad Tree} & \multicolumn{2}{c}{\quad Loop} & \multicolumn{2}{c}{$\delta_{\rm exp}$} & \multicolumn{2}{c}{$\delta_{\rm th}$}\\ 
\midrule
$\mcha{1}$ & \quad$149$\quad & $6$ & \multicolumn{2}{c}{\quad$-$} & \quad$0$ & $1\ (0.2)$ & \multicolumn{2}{c}{$-$} & \quad$139$ & $3$ & \multicolumn{2}{c}{\quad$-$} & \quad\,\,$0$ & $1$ & $-$\\ 
$\mcha{2}$ & $292$ & $3$ & \multicolumn{2}{c}{\quad$-$} & $0$ & $5\ (2)$ & \multicolumn{2}{c}{$-$} & $266$ & $2$ & \multicolumn{2}{c}{\quad$-$} & $0$ & $5$ & $-$\\ 
$\mneu{1}$ & $106$ & $9$ & \multicolumn{2}{c}{\quad$-$} & $0$ & $2$ & \multicolumn{2}{c}{$-$} & $92$ & $8$ & \multicolumn{2}{c}{\quad$-$} & $0$ & $2$ & $-$\\
$\mneu{2}$ & $164$ & $0$ & $2$ & $0$ & $0$ & $5\ (1)$ & $0$ & $5$ & $148$ & $5$ & $\qquad2$ & $4$ & $0$ & $5$ & $0.5$\\ 
$\mneu{3}$ & $188$ & $6$ & $-1$ & $5$ & $0$ & $5\ (1)$ & $0$ & $5$ & $189$ & $7$ & $-7$ & $3$ & $0$ & $5$ & $0.5$\\ 
$\sigma(\cha^+_1\cha^-_1)^{350/400}_{(-0.8,0.6)}$ & $2347$ & $5$ & $-291$ & $3$ & $8$ & $7$ & $2$ & $0$
 & $709$ & $7$ & $-85$ & $1$ & $0$ & $7$ & $-$ \\
$\sigma(\cha^+_1\cha^-_1)^{350/400}_{(0.8,-0.6)}$ & $224$ & $4 $ & $7$ & $6 $ & $2$ & $7$ & $0$ & $5$
& $129$ & $8 $ & $20$ & $0 $ & $0$ & $3$ & $-$ \\
$\sigma(\cha^+_1\cha^-_1)^{500}_{(-0.8,0.6)}$ & $1450$ & $6 $ & $-24$ & $4$ & $6$ & $7$  & $2$ & $0$
& $560$ & $0 $ & $ -70$ & $1$ & $0$ & $7$ & $-$ \\
$\sigma(\cha^+_1\cha^-_1)^{500}_{(0.8,-0.6)}$ & $154$ & $8  $ & $ 12$ & $7 $ & $2$ & $0$ & $0$ & $5$
& $97$ & $1  $ & $ 16$ & $4 $ & $0$ & $3$ & $-$ \\
$A_{FB}^{350/400}(\%)$ & $-2$&2& 6&8&0&8&0&1& 24&7&$-2$&8& 1&4 & $0.1$\\
$A_{FB}^{500}(\%)$ &$-2$&6 &5&3&1&0&0&1 &39&2&$-5$&8 & 1&5 & $0.1$\\  
\bottomrule
\end{tabular}
\end{center}
\caption{Observables (masses in GeV, cross sections in fb) used as input for the fit in S1 (left) 
and S2 (right), tree-level values and loop corrections are specified along with the expected experimental and theoretical errors ($\delta_{\rm exp}$ and $\delta_{\rm th}$). Here the superscript on 
$\sigma$ and $A_{FB}$, i.e. $350\,/\,400$ (for $\rm S1\,/\,S2$) and 500, denotes $\sqrt{s}$ in 
GeV, and the subscript on $\sigma$ denotes the beam polarisation $(\mathcal{P}(e^-),\mathcal{P}(e^+))$. 
The central value of the theoretical prediction, $\mathcal{B}(b\to s\gamma)= 3.3\cdot 10^{-4}$ GeV for 
S1, and $\mathcal{B}(b\to s\gamma)=2.7\cdot 10^{-4}$, $\Delta (g_\mu -2)/2=2.4\cdot 10^{-9}$ and $m_h=125$ 
GeV for S2 are also included in the fit. Errors in brackets are for masses obtained from the continuum. 
See text for details of error estimation.}\label{tab:inputsc1} 
\end{table}
The results for S1, given in tab.~\ref{tab:ressc1}, show the fit to the 8 MSSM parameters: $M_1$, $M_2$, $\mu$, $\tan\beta$,
$m_{\tilde{\nu}}$, $\cos\theta_{\tilde{t}}$, $m_{\tilde{t}_1}$, and
$m_{\tilde{t}_2}$.
We find that the gaugino and higgsino mass parameters are determined with an 
accuracy better than 1\%, while $\tan\beta$ is determined with an accuracy of 
$5\%$, and  2-3\% for the sneutrino mass.
The limited access to the stop sector (tab.~\ref{tab:ressc1}) could nevertheless lead to
hints allowing a well-targeted search at the LHC. 
In tab.~\ref{tab:ressc1} we also compare the fit results obtained using masses
of the charginos and neutralinos from threshold scans to those obtained using masses from the
continuum. For the latter, the fit
quality deteriorates,
clearly indicating the need to measure these masses via
threshold scans.
For S2, the heavy Higgs boson mass is added to and the sneutrino mass is removed from the set of fit parameters. 
The results seen in tab.~\ref{tab:ressc1} show that the electroweakino parameters are precisely determined, 
and that the fit is sensitive to $m_{\tilde{t}_2}$, with an accuracy better than $20\%$. In
addition, an upper limit on the mass of the heavy Higgs boson can be placed at 1000~GeV,
at the 2$\sigma$ level. 
\begin{table}[tb!]
\renewcommand{\arraystretch}{1.1}
%\vspace{1cm}
\begin{center}
\begin{tabular}{lr@{}l@{ }l r@{}l@{ }l r@{}l@{ }l }\toprule
&\multicolumn{6}{c}{\footnotesize S1}&\multicolumn{3}{c}{\footnotesize S2}\\
\vspace{-0.9cm} \\
&\multicolumn{6}{c}{\downbracefill}&\multicolumn{3}{c}{\downbracefill}\\
\T 
Parameter & \multicolumn{3}{c}{Threshold fit} & \multicolumn{3}{c}{Continuum fit} & \multicolumn{3}{c}{Threshold fit}\\
\midrule
$M_1$ &  $125 $ & $\pm 0.3 $ & $ (\pm 0.7) $ & $125 $ & $\pm 0.6 $ & $ (\pm 1.2)$ &   $106$ &$\pm 0.3$ & $(\pm 0.5) $  \\  
$M_2$ & $250 $ & $\pm 0.6 $ & $ (\pm 1.3) $  & $250 $ & $\pm 1.6 $ & $ (\pm 3)$ & $212$ &$\pm 0.5 $ & $ (\pm 1.0) $   \\  
$\mu$ &  $180 $ & $\pm 0.4 $ & $ (\pm 0.8) $ & $180 $ & $\pm 0.7 $ & $ (\pm 1.3)$ &  $180$ &$\pm 0.4 $ & $ (\pm 0.9) $  \\  
$\tan\beta$ & $10 $ & $ \pm 0.5 $ & $ (\pm 1) $  & $10 $ & $\pm 1.3 $ & $ (\pm 2.6) $ &  $12$ &$\pm 0.3 $ & $ (\pm 0.7) $ \\  
$m_{\tilde{\nu}}$ & $1500 $ & $\pm 24 $ & $ (^{+60}_{-40}) $  & $1500 $ & $\pm 20 $ & $ (\pm 40) $&\multicolumn{3}{c}{$-$}\\
$\cos\theta_{\tilde{t}}$ & $0.15$&$^{+0.08}_{-0.06} $ & $ (^{+0.16}_{-0.09}) $ & $0 $ & $\pm 0.15 $ & $ (^{+0.4}_{-0.3}) $ & \multicolumn{3}{c}{$-$} \\ 
$m_{\tilde{t}_1}$ & $400 $ & $^{+180}_{-120} $ & $ (^{\textrm{at limit}}_{\textrm{at
limit}}) $  & \multicolumn{3}{c}{$-$} &  $430$&$^{+200}_{-130} $ & $ (^{+300}_{-400}) $  \\  
$m_{\tilde{t}_2}$ & $800$ & $^{+300}_{-170} $ & $ (^{+1000}_{-290}) $  &
$800$ & $^{+350}_{-220} $ & $  (^{\textrm{at limit}}_{\textrm{at
limit}}$) & $1520$&$^{+200}_{-300} $ & $ (^{+300}_{-400}) $  \\
$m_{A^0}$ &\multicolumn{3}{c}{$-$}&\multicolumn{3}{c}{$-$}& \multicolumn{2}{r}{$<650$}  & $ (<1000) $ \\ \bottomrule
\end{tabular}\hspace{.8cm}
\end{center}
\caption{Fit results (masses in GeV) for S1 (left) and S2 (right), for masses obtained from threshold scans
(threshold fit) 
and from the continuum (continuum fit). Numbers in brackets denote $2\sigma$
errors.\label{tab:ressc1}}
\end{table}

\section{Conclusion}\label{sec:conc}
We have extended previous analyses to extract fundamental MSSM parameters via chargino
production at the LC, by incorporating loop corrections. The polarised cross-sections and forward backward asymmetry for chargino
production as well as the $\cha_1^{\pm},\cha_2^{\pm}$
and $\neu{1}, \neu{2}, \neu{3}$ masses, are calculated at NLO in an on-shell scheme.
When  $M_1$, $M_2$ and $\mu$ are
light, we determine them at $\mathcal{O}(\%)$, and $\tan\beta$ to $<5\%$.
Masses obtained from the continuum
as opposed to threshold scans result in doubled uncertainties on the fundamental parameters.
For our 125 GeV Higgs mass compatible scenario we obtain
an accuracy of better than $20\%$ on the mass of the heavy, left-handed stop.
In summary, incorporating NLO corrections is shown to be required for the
precise determination of the fundamental electroweakino 
parameters at the LC, and to provide sensitivity to the parameters
describing particles contributing via loops.
\bibliographystyle{unsrt}
\bibliography{BKMRW}
\end{document}